\documentclass[runningheads]{llncs}
\usepackage[T1]{fontenc}
\usepackage{booktabs}
\usepackage[misc]{ifsym}
\usepackage{cite}
\usepackage{amsmath,amssymb,amsfonts}
\usepackage{algorithmic}
\usepackage{graphicx}
\usepackage{textcomp}
\usepackage{xcolor}
\usepackage[hidelinks]{hyperref}
\usepackage{cleveref}
\usepackage{siunitx}
\usepackage{subfig}
\usepackage{float}

\usepackage{mwe}

\begin{document}

\title{MultiVic: A Time-Predictable RISC-V Multi-Core Processor Optimized for Neural Network Inference \thanks{This work was supported by the German Federal Ministry of Research, Technology and Space (BMFTR) within the project "MANNHEIM-CeCaS," funding number 16ME0818.}}

\titlerunning{Time-Predictable RISC-V Multi-core Processor for Neural Network Inference}

\author{Maximilian Kirschner\inst{1,2} \and Konstantin Dudzik\inst{1,2} \and Ben Krusekamp\inst{1,2} \and J\"urgen Becker\inst{1,2}}

\authorrunning{M. Kirschner et al.}

\institute{FZI Research Center for Information Technology, Karlsruhe, Germany
\and
Karlsruhe Institute of Technology, Karlsruhe, Germany\\\email{\{kirschner, dudzik, krusekamp, juergen.becker\}@fzi.de}}

\maketitle              

\begin{abstract}
Real-time systems, particularly those used in domains like automated driving, are increasingly adopting neural networks. From this trend arises the need for high-performance hardware exhibiting predictable timing behavior. While state-of-the-art real-time hardware often suffers from limited memory and compute resources, modern AI accelerators typically lack the crucial predictability due to memory interference.

We present a new hardware architecture to bridge this gap between performance and predictability. The architecture features a multi-core vector processor with predictable cores, each equipped with local scratchpad memories. A central management core orchestrates access to shared external memory following a statically determined schedule. 

To evaluate the proposed hardware architecture, we analyze different variants of our parameterized design. We compare these variants to a baseline architecture consisting of a single-core vector processor with large vector registers. We find that configurations with a larger number of smaller cores achieve better performance due to increased effective memory bandwidth and higher clock frequencies.  Crucially for real-time systems, execution time fluctuation remains very low, demonstrating the platform's time predictability.

\end{abstract}

\section{Introduction}
Neural networks are being used in an increasing number of real-time systems, especially in automotive or aviation applications. This results in demanding requirements for the processing hardware used. On one hand, powerful and highly parallel computing units and large amounts of fast memory are required to execute typical deep neural networks. On the other hand, it is necessary to provide guarantees for the time behavior of the execution. The hardware must therefore have time-predictable properties.

Current processor and accelerator platforms each lack one of these features. GPUs and modern accelerator architectures offer high performance, but are not time-predictable due to their execution model and interference in shared memory~\cite{gpu_interference, soc_interference}.
Multi-core systems developed for real-time applications typically rely on time-division multiple access (TDMA) on the system bus or Network on Chip (NoC) to avoid memory interference~\cite{survey}. 
But, this limits the usable memory bandwidth.
Vector processors are a promising approach to increase the compute performance of real-time systems without compromising predictability \cite{vicuna}. However, their scalability in terms of vector length and width of the computing units is limited, necessitating the use of multiple processors for typical ML inference applications. 

The main contribution of this work is therefore a novel multi-core architecture optimized for time-predictable neural network inference. We present the parameterized architecture concept, implementation details and a extensive evaluation of different processor configurations. The implementation of the hardware architecture, usable on FPGAs, will be provided as open source.


The remainder of this work is structured as follows:
\Cref{sec:related-work} gives a brief overview over the related work in the field of time-predictable processors and embedded machine learning acceleration. Our novel multi-core architecture is then presented in \Cref{sec:concept}, with a discussion about design considerations regarding performance and time-predictability. \Cref{sec:implementation} shows details about the hardware implementation, explains the deployment of software on the platform and describes the benchmark implementation. In \Cref{sec:evaluation} various configurations of the presented architecture are evaluated and compared regarding performance and resource consumption against baseline hardware with a single-core vector processor. We also describe potential improvements  that can be derived from the evaluation results. \Cref{sec:conclusion} concludes and summarizes the contributions and findings.

\section{Related Work}\label{sec:related-work}
Accelerating neural networks (NNs) often involves domain-specific accelerators, which offer high efficiency but lack flexibility, requiring extensive redesign for new models. Our approach, in contrast, utilizes a programmable general-purpose architecture capable of executing arbitrary models.

RISC-V vector processors are popular for embedded machine learning acceleration due to their efficient parallelism. While examples like Spatz \cite{perotti2025} and Ara2 \cite{perotti2024} exist, they can introduce timing anomalies that complicate worst-case execution time (WCET) analysis. Our architecture incorporates the Vicuna vector co-processor \cite{vicuna}, which is specifically designed to be free of such anomalies, enabling precise timing analysis crucial for real-time systems.

Multi-core clusters are another common approach for NN acceleration in embedded systems, exemplified by platforms like Kalray MPPA-256 \cite{deDinechin2013}, Mr. Wolf \cite{pullini2019}, and MemPool \cite{riedel2023}. These platforms typically use shared memory for data exchange, but this often leads to unpredictable memory interference, making them unsuitable for hard real-time applications.

However, there are also multi-core platforms designed with a focus on predictability. These platforms implement the concept of Precision Timed (PRET) Machines \cite{edwards2007}. The platforms use time-division multiple access (TDMA) and distributed
memories to avoid interference. The Merasa processor \cite{paolieri2013} uses TDMA on the bus for this purpose. Other platforms
utilize TDMA NoCs. Examples include the T-CREST platform \cite{schoeberl2015}, parMERASA \cite{ungerer2016} and InterPRET \cite{jellum2023}.

However, many PRET architectures struggle with efficient external memory integration and management, limiting their performance for data-intensive NNs. Furthermore, TDMA memory arbitration restricts each core's access to available memory bandwidth. Our approach, however, relies on explicitly scheduled data transfers to local memories, which allows for independent, concurrent execution and increases effective memory bandwidth.

In addition, we incorporate a management core with a DMA for central orchestration, executing the data transfers as scheduled at compile time. This is feasible for the execution of classic feed-forward neural networks due to the deterministic data flow.

In \cite{kirschner2024} we proposed the concept of a heterogeneous multi-core system with a dedicated management core and specialized worker cores for neural network acceleration in real-time systems. This work builds upon said concept by providing a detailed hardware implementation and evaluation of different configurations.

\section{Proposed Hardware Architecture}\label{sec:concept}
In order to design a predictable processor capable of executing practically relevant machine learning applications, several requirements need to be fulfilled.
Efficient inference of neural networks, especially convolutional neural networks (CNNs), necessitates a high degree of parallelization and large memory capacities.
Additionally, the architecture should offer general-purpose compute capabilities to be flexible regarding supported layer types and network architectures. 
When designing for predictable execution, one of the most important requirements is freedom from interference in the memory subsystem. We therefore place particular focus on this.

In state-of-the-art accelerator systems, many-core architectures and vector processors are dominant approaches, offering the necessary degree of parallelization. In this work, we propose a multi-core vector processor that integrates both concepts, thereby providing two dimensions of parallelization. A key feature of the concept is the predictable execution of each core on its own local scratchpad memories. These are split into an instruction and a data memory. Each core can only access its local scratchpad memories. The absence of shared resources, such as global memories or peripherals, that could be accessed by multiple cores, guarantees freedom from interference by design.

The execution of neural networks requires communication between cores. A dedicated management core coordinates memory transactions between the scratchpads, while also enabling communication to a global DRAM, which is employed for storing weights and intermediate activations. To offload copy operations from the management core, a DMA peripheral is utilized. Dual-port SRAMs are used for the scratchpads to enable simultaneous access of the worker core and management core without interference.

\begin{figure}[h]
    \centering
    \includegraphics[width=0.65\linewidth]{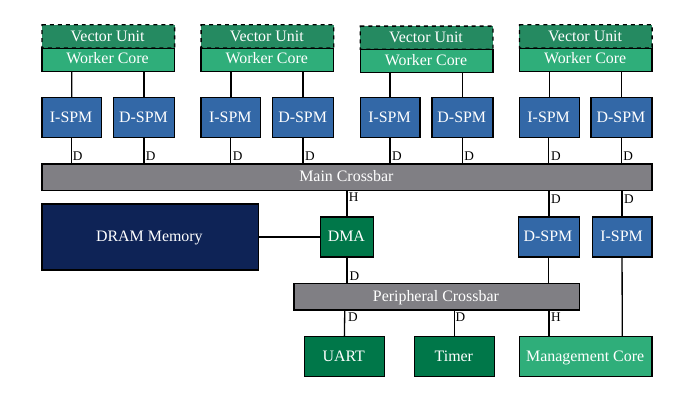}
    \caption{The proposed multi-core processor architecture, here with four cores}
    \label{img:multicore-architecture}
\end{figure}

\autoref{img:multicore-architecture} shows an overview of the architecture. It can be seen, that each crossbar has only one host and multiple devices attached. This means that only one component, the management core or the DMA (which is controlled by the management core) can initiate data transfers. This guarantees freedom from interference.

However, this kind of central orchestration comes with scheduling challenges. But given the deterministic data flow of the targeted neural networks, the scheduling can be solved statically at compile time. This results in a limitation for the supported neural networks: The data flow must be deterministic and must not depend on the input data to enable scheduling of memory transactions at compile time. This is not a problem for standard CNNs. However, if dynamic CNNs are to be used or sparsity is to be exploited, additional assumptions must be made during scheduling. Initial approaches for the scheduling of computations and memory transactions are described in \cite{kirschner2024}.

\subsection{Discussion}
The hardware architecture delivers compute resources with two dimensions of parallelism and extensive memory connectivity, essential for neural network execution. This design enables efficient utilization of compute units and memory bandwidth through effective compile-time scheduling.

Each worker core is implemented as a full-fledged RISC-V vector processor, offering significant flexibility. This design not only facilitates efficient convolution computations but also enables the mapping of diverse layer types onto the worker cores. This ensures that performance is not limited by the management core.

The architecture inherently maintains time-predictability. A cornerstone of this predictability is timing compositionality, a property present when the hardware architecture is free from timing anomalies. As established by \cite{timing-composition}, timing compositionality permits a safe decomposition of timing analysis into individual components. This allows for the derivation of a global Worst-Case Execution Time (WCET) based on localized worst-case behavior. \cite{vicuna} further demonstrates this property for the Vicuna co-processor when integrated with the in-order pipeline of the Ibex core.

Consequently, each vector core within the presented architecture is predictable when operating independently. This independence is rigorously guaranteed because each core operates exclusively on its own dedicated memory, eliminating shared resources and inter-core interference. While the management core can access the scratchpads of other cores, these accesses do not introduce interference.  This is because the utilized SRAMs feature two independent ports and consistently respond within a single clock cycle.

The only potential factor influencing a vector core's local execution time is the timely availability of required data in its data scratchpad. However, given the management core's static schedule, both data availability and computation start times are predictable at compile time. The DRAM is the sole component introducing timing variability into the system due to its variable memory access times. Therefore, a worst-case model for the DRAM must be assumed and integrated into the static schedule calculation. This ensures that DRAM accesses never exceed their allocated duration within the schedule.

\section{Implementation}\label{sec:implementation}
\subsection{Hardware Design}
The architecture is implemented in SystemVerilog for evaluation on an FPGA. The source code is published under a permissive open-source license\footnote{\url{https://github.com/MultiVic}}. 
All cores in the system are based on the Ibex \cite{ibex} RISC-V core, with the Vicuna \cite{vicuna} vector co-processor attached in the worker cores. The interconnects are based on the TileLink Uncached-Lightweight standard, implemented using the OpenTitan \cite{OpenTitan} TileLink generator. For the DMA, a modified version of the OpenTitan DMA is utilized. The \qty{4.5}{\giga\byte} DDR4 memory, which is available on the FPGA board, is employed as the main memory, in conjunction with the Xilinx memory controller.

We can configure the implementation using several parameters to create and test various processor setups. The Vicuna core itself has multiple options for its vector pipeline, most notably the vector register size and compute unit width.
For our multi-core implementation, the number of cores isn't fixed, allowing us to generate different versions. We can also configure the size of the scratchpads. In the subsequent section we compare configurations using a few large cores and memories against those with many smaller cores and memories.

\subsection{Deployment}
In future work, we plan to develop an MLIR-based compiler flow for our hardware architecture, which will enable the deployment of entire neural networks on the platform, as described in \cite{kirschner2024}. This compiler will automate the mapping of computations to the cores and the scheduling of DMA transfers. However, this development is outside the scope of this paper. To evaluate the hardware, we manually implement the fundamental mechanisms for the loading of instructions and data to the scratchpads. On top of this a matrix multiplication benchmark is implemented.

It is important to note that the local memories have different addresses from the perspective of the worker cores compared to the global view of the management core and DMA. To facilitate the development of our benchmarks, we created code generation scripts. These scripts compile the programs for each individual core and generate a linker script that links the object files into separate segments within a single large binary. This binary is first loaded into the memory of the management core. From there, it is distributed to the instruction scratchpads of the worker cores. 

To execute a given schedule, the management core has access to a timer for time-triggered schedules. Additionally, the management core can retrieve the status of the worker cores. For this purpose, a dedicated area is reserved in each worker core's data scratchpad for the exchange of status and control information. On the worker cores, a minimal runtime environment is executed. This runtime sets a status bit to indicate whether an execution is in progress and, when a new execution is to be started, jumps to the provided function pointer. After the function completes, the program returns to the runtime. The management core can see if an execution is currently in progress and pass function pointers to indicate the next function to be executed. 

\subsection{Matrix Multiplication Benchmark}
Convolutions and fully connected layers are the most common and computationally intensive layer types in CNNs. These layer types are usually implemented using matrix multiplictaions.
To evaluate the performance of our multi-core architecture with distributed memories, we therefore implement a matrix multiplication benchmark. The calculations are distributed to the available cores. Runs are always started when previous computations have been completed. In this way, the maximum throughput can be determined. For more complex programs such as entire networks, time-triggered execution is preferable to facilitate timing analyses.

In the \textit{matmul} benchmark, the computation \(\mathbf{A}\cdot\mathbf{B}=\mathbf{C}\) is performed. In our implementation, the matrices are square, with \(\mathbf{A},\mathbf{B},\mathbf{C} \in \mathbb{R}^{N \times N}\) and \({N=1024}\).
To distribute the computations across the worker cores, matrix \(\mathbf{B}\) is partitioned into blocks of width \(B\). These blocks are then distributed to the data scratchpads of the worker cores, where they remain for as long as possible to minimize memory transfers to the main memory.

Iteratively, the rows of \(\mathbf{A}\) are loaded and processed on all cores. This process generates fragments of the rows of \(\mathbf{C}\), which are subsequently written back to the main memory.
To leverage the vector units, within an inner loop, the rows of \(\mathbf{A}\) and the columns of \(\mathbf{B}\) are decomposed into vectors with a length equal to the vector register size. This enables vectorized execution of the multiplications and vectorized accumulation of the products to calculate individual elements of \(\mathbf{C}\).
%
%
\section{Evaluation}\label{sec:evaluation}
The evaluation aims to investigate the performance of the platform in comparison to a single vector processor. The research question we try to answer is, if a time-predictable multi-core architecture has a higher performance than a single large predictable vector processor, despite the overhead of explicit data movement between the cores.
It also seeks to determine the configuration that achieves optimal performance. Furthermore, the study will examine the scalability of the architecture and the limitations of scalability.

To facilitate a comparison with a single-core platform, we define a baseline architecture as a reference, depicted in Figure 3. This baseline consists of a single Ibex core integrated with a Vicuna vector co-processor, connected to a 64 KiB instruction scratchpad and a 1 MiB data scratchpad. Additionally, this core has access to the same peripherals as the multi-core system, including DMA with an interface to the DDR4 main memory. For the Vicuna co-processor itself, we utilize the \textit{Fast} configuration as described in the original paper by Platzer and Puschner \cite{vicuna}, with details provided in \autoref{tab:baseline-max-frequency}.

\begin{table*}
    \centering
    \caption{Baseline System Configurations with Maximum Achievable Clock Frequencies}
    \label{tab:baseline-max-frequency}
    \begin{tabular}{|c|c|c|c|}
        \hline
        \textbf{Configuration} & \textbf{Vector Register Width} & \textbf{Multiplier Width} & $\mathbf{F_{max}}$ \textbf{(VCU128)} \\ \hline
        Small		& 128					& 32				 & \qty{179}{\mega\hertz} \\ \hline
        Medium		& 512					& 128				 & \qty{177}{\mega\hertz} \\ \hline
        Fast		& 2048					& 1024				 & \qty{149}{\mega\hertz} \\ \hline
    \end{tabular}
\end{table*}

The evaluation explores four distinct configurations of the multi-core architecture, as summarized in \autoref{tab:multicore-configurations}. These configurations primarily differ in the number of worker cores. The size of the vector unit within the worker cores is inversely scaled with the core count. For example, the dual-core variant features half the vector register length and half the width of the multiplication unit compared to the \textit{Fast} configuration.

\begin{figure}[H]
    \centering
    \includegraphics[width=0.55\linewidth]{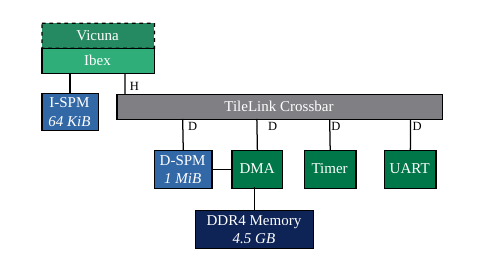}
    \caption{Baseline architecture with a single core and similar peripherals}
    \label{img:baseline-architecture}
\end{figure}

The data scratchpad size is also scaled inversely proportional to the number of cores. While the baseline hardware is equipped with a \qty{1}{\mebi\byte} data scratchpad, the multi-core variants have \qty{512}{\kibi\byte}, \qty{256}{\kibi\byte}, \qty{128}{\kibi\byte}, and \qty{64}{\kibi\byte}, per core respectively. The instruction scratchpad size for the worker cores remains constant at \qty{16}{\kibi\byte}. The instruction and data scratchpads for the management core are consistently \qty{64}{\kibi\byte} each.

\begin{table*}
    \centering
    \caption{Configuration Variants of the Multi-core System Used in Evaluation}
    \label{tab:multicore-configurations}
    \resizebox{\textwidth}{!}{
    \begin{tabular}{|c|c|c|c|c|c|}
        \hline
        \textbf{Configuration} & \textbf{\# Worker Cores} & \textbf{Vector Register Width} & \textbf{Multiplier Width}  & \textbf{Data SPM Size} & $\mathbf{F_{max}}$ \textbf{(VCU128)} \\ \hline
        Dual		& 2					& 1024				& 512					& \qty{512}{\kibi\byte} & \qty{168}{\mega\hertz} \\ \hline
        Quad		& 4					& 512				& 256					& \qty{256}{\kibi\byte} & \qty{169}{\mega\hertz} \\ \hline
        Octa		& 8					& 256				& 128					& \qty{128}{\kibi\byte} & \qty{168}{\mega\hertz} \\ \hline
        Hexadeca	& 16				& 128				& 64					& \qty{64}{\kibi\byte} & \qty{118}{\mega\hertz} \\ \hline
    \end{tabular}}
\end{table*}

\subsection{Performance}
\begin{figure}
    \centering
    \includegraphics[width=0.65\linewidth]{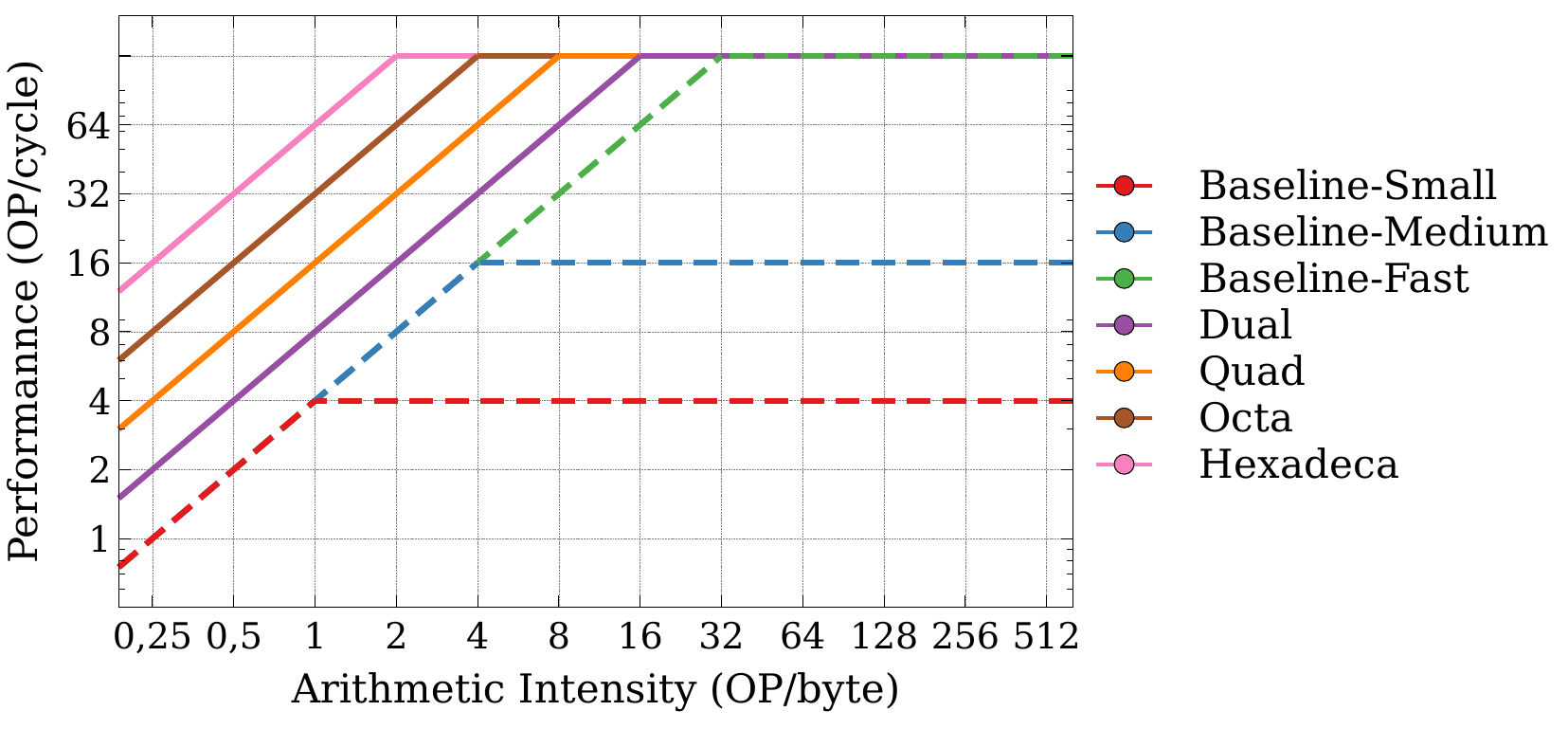}
    \caption{Roofline plot comparing theoretical performance}
    \label{img:roofline}
\end{figure}

Compared to a single-core vector processor, the multi-core architecture possesses a higher total bandwidth to the fast scratchpad memory. Each core features its own dedicated connection to the SRAM. Figure \ref{img:roofline} presents a roofline plot illustrating the theoretical performance for the individual configurations.
The multi-core variants achieve the same computational performance as the \textit{Fast} configuration but effectively shift the memory boundary. Consequently, the multi-core architecture particularly benefits data-intensive algorithms exhibiting high data reuse, thus minimizing the requirement for DMA transfers.

The multi-core architecture utilizing smaller vector processors offers a further advantage compared to a single large vector processor: its design has shorter critical paths, enabling the achievement of higher clock frequencies.
To evaluate this, the maximum achievable clock frequencies on the VCU128 Virtex Ultrascale+ FPGA board are iteratively determined using Xilinx Vivado. For this process, the clock constraints are progressively tightened in each synthesis run based on the worst negative slack reported in the preceding run. This iterative tightening continues until Vivado is unable to achieve timing closure. The determined maximum frequencies are presented in \autoref{tab:baseline-max-frequency} and \autoref{tab:multicore-configurations}.

It can be observed that the multi-core variants can achieve significantly higher clock frequencies than the \textit{Fast} configuration, which is comparable in theoretical performance. However, congestion in the FPGA routers emerges at 16 cores, primarily due to the necessity of routing connections from 34 scratchpads (16 worker cores + management core) to the DMA. This congestion leads to a notable decrease in the achievable clock frequency, thus marking a limitation in scalability at this point.

\subsubsection{Matrix Multiplication}
To evaluate the performance of the hardware platform, the \textit{matmul} benchmark was executed 100 times on each of the configuration variants. The hardware was operated at a clock frequency of 100 MHz, and execution time was measured in clock cycles using the timer peripheral.

\begin{figure}
    \centering
    \includegraphics[width=0.65\linewidth]{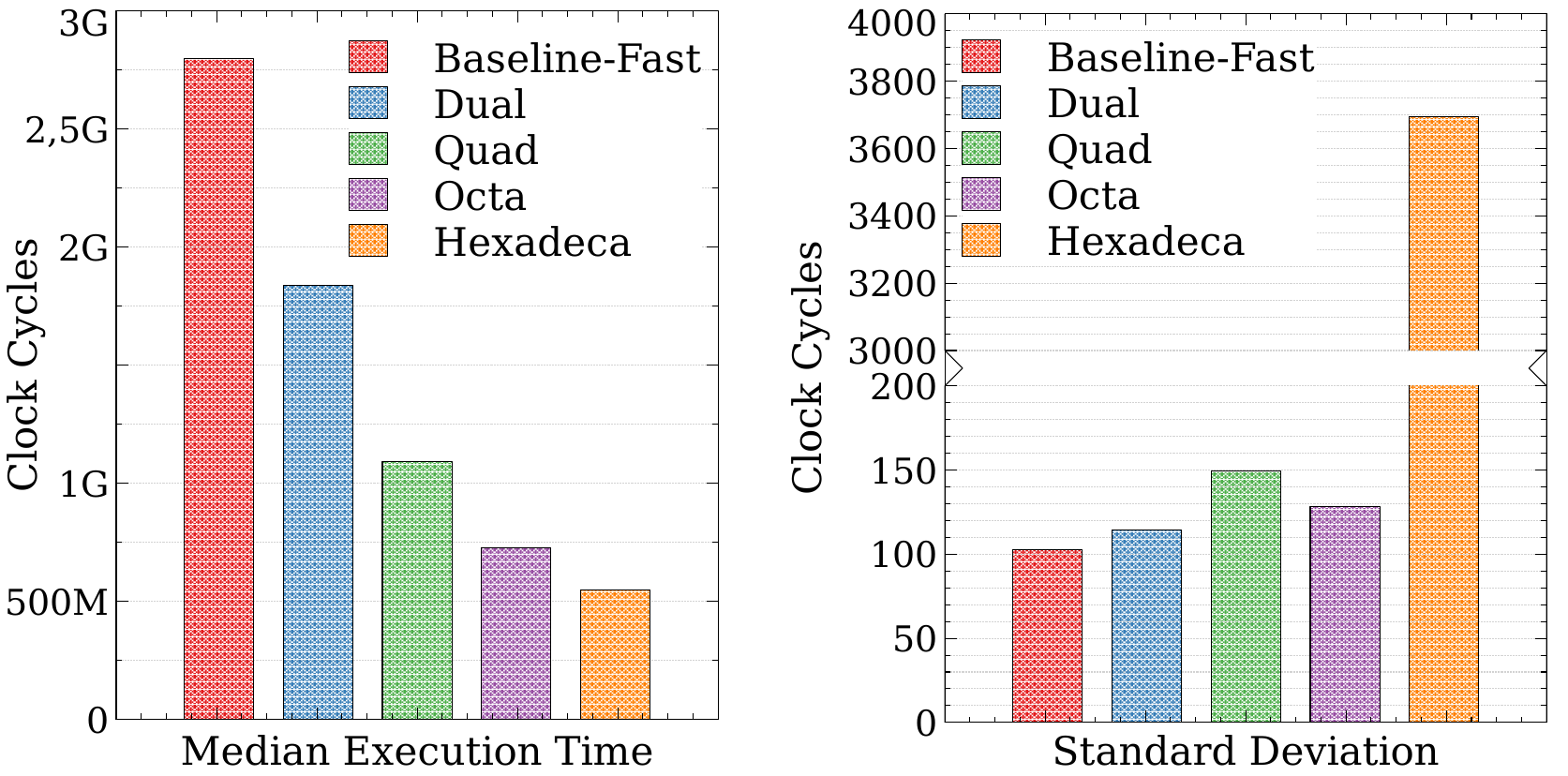}
    \caption{Median execution time and standard deviation on the matmul benchmark}
    \label{img:matmul-results}
\end{figure}

The measurement results are presented in Figure \ref{img:matmul-results}. The left bar plot shows the median execution time for each configuration, while the right bar plot displays the standard deviation of the execution times for each configuration.
It can be observed that the median execution times decrease with an increasing number of cores. This is attributed to the increased memory bandwidth associated with a higher core count. The parallelism can be effectively exploited, allowing multiple worker cores to simultaneously load their vector registers.
However, performance does not scale indefinitely. The improvement in execution times between 8 and 16 cores is only minimal. Here, the duration of the DMA transfers and the calculations in shorter vector registers outweigh the gained scratchpad bandwidth.

The variability of the execution times is very low relative to the median execution time.
These fluctuations come from the fluctuating access times of the DDR4 memory.
The variability increases with the number of cores. This is because the matrix is broken down into smaller and smaller blocks. This increases the number of DDR4 accesses, which in turn increases the execution time variability.

In summary, the \textit{Octa} configuration is optimal for the \textit{matmul} benchmark. In median the \textit{Octa} configuration needs $\qty{728548804}{cycles}$ @ $\qty{168}{\mega\hertz} \approx \qty{4.33}{\second}$, while the \textit{Hexadeca} configuration needs $\qty{548343601}{cycles}$ @ $\qty{118}{\mega\hertz} \approx \qty{4.65}{\second}$. The \textit{Octa} configuration additionally delivers a very low execution time variance. 

%

\subsection{Resource Consumption}
\begin{figure}[H]
    \centering
    \subfloat[Comparison of different architecture variants\label{img:resources-configurations}]{
        \includegraphics[width=0.45\linewidth]{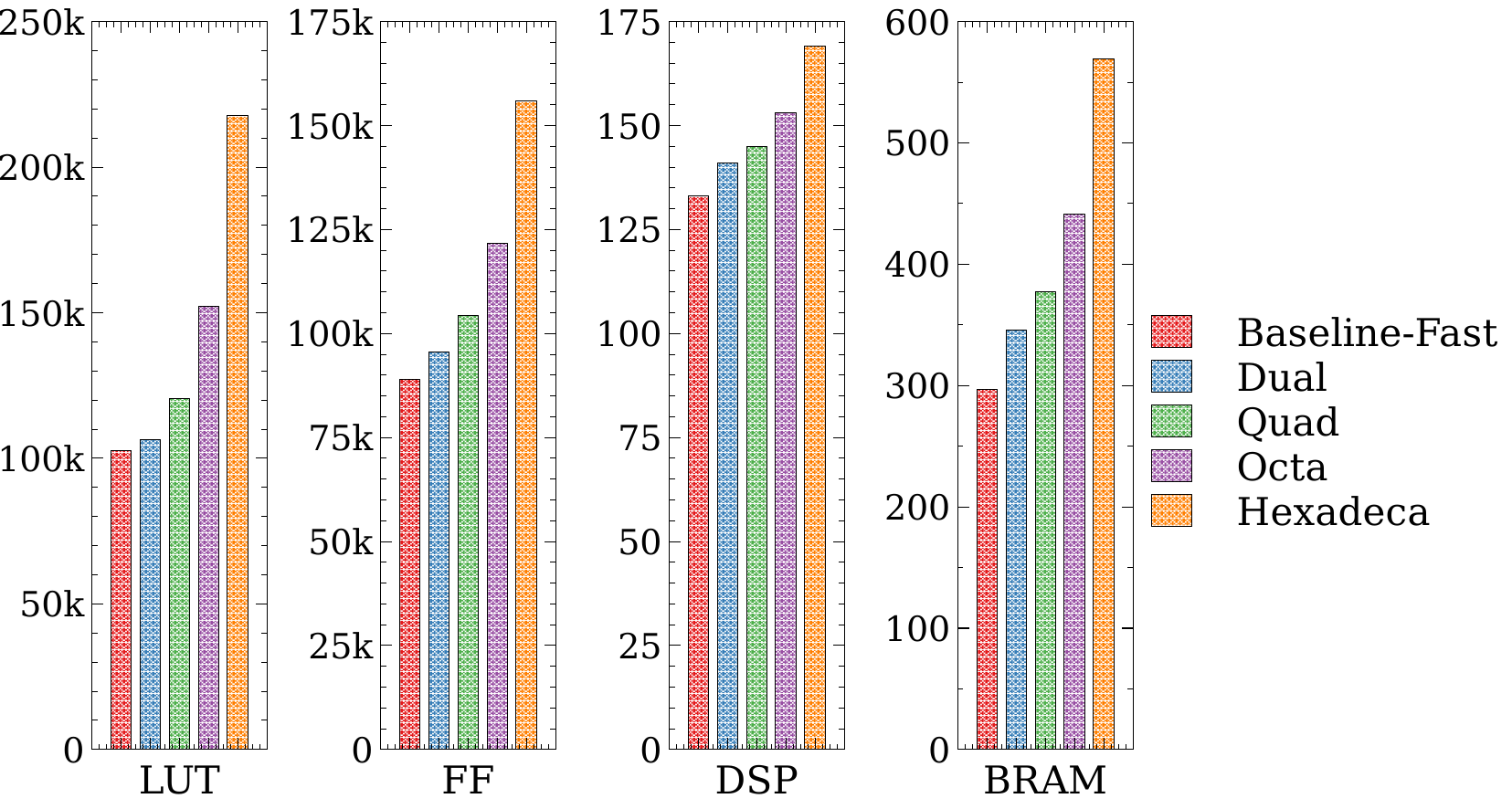}
    }
    \hfil
    \subfloat[Components in a dual core system\label{img:resources-components}]{
        \includegraphics[width=0.45\linewidth]{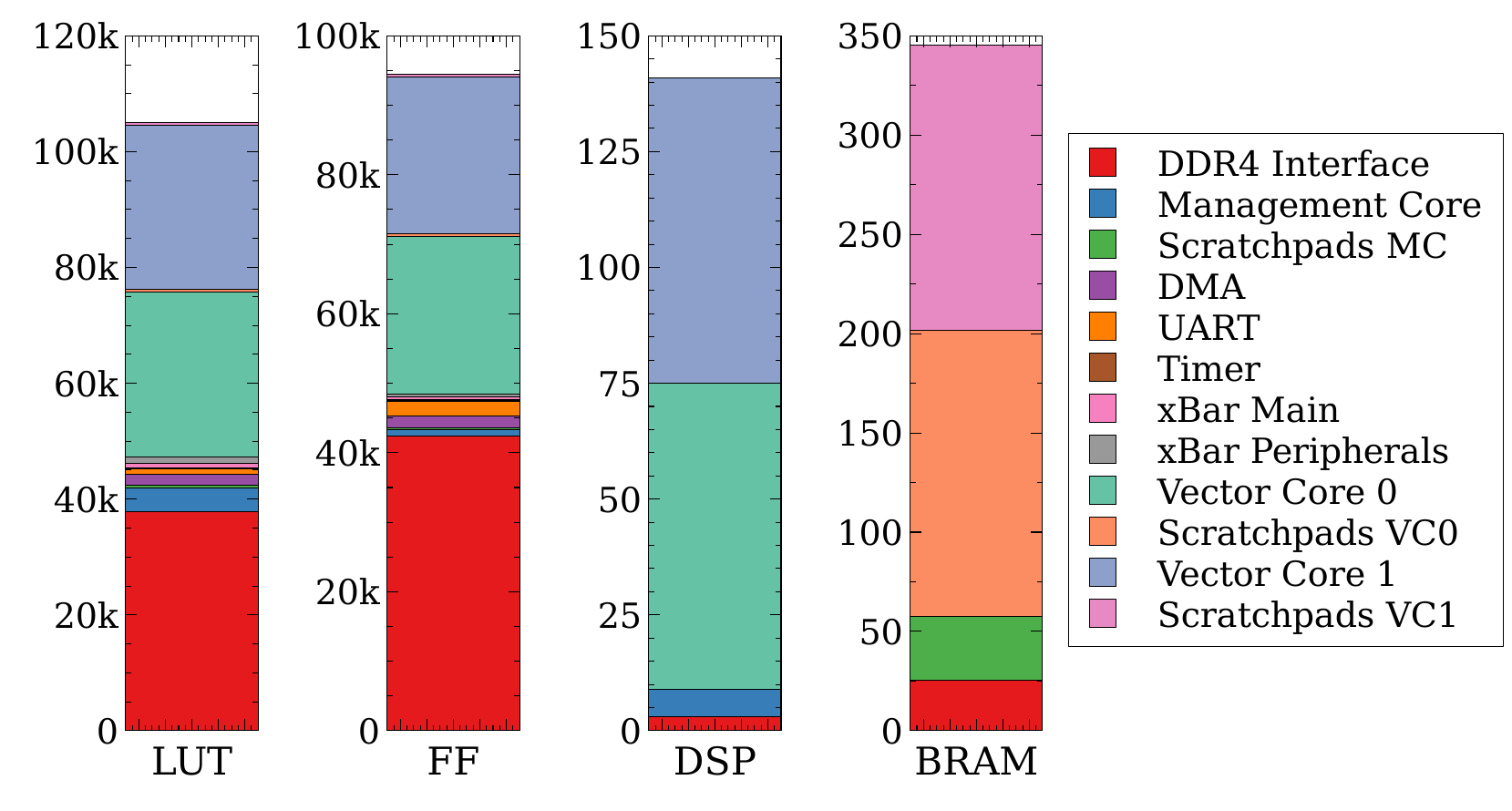}
    }
    \caption{FPGA resource consumption on VCU128}
\end{figure}

In addition to the raw performance, the FPGA resource utilization of the different configurations is evaluated. For this purpose, the values reported by Xilinx Vivado during the synthesis run at the maximum achievable clock frequency are used. Figure \ref{img:resources-configurations} illustrates the utilized look-up tables, flip-flops, block RAM, and DSP cells, for the different configurations.

We observe that the overall resource consumption increases with the number of cores. This is primarily because each additional core adds an instruction scratchpad and the complete logic of an Ibex core. However, by reducing the size of the vector units and data scratchpads, we still observe good scalability. The count of DSP cells, in particular, indicates that employing many small vector processors does not require significantly more logic than using a few large ones.

To better assess the scalability of the architecture, the resource consumption of individual system components is also considered. This detailed resource consumption is depicted in Figure \ref{img:resources-components} for the \textit{Dual} configuration.
It can be observed that the management core, with its local memories and peripherals, accounts for only a very small portion of the total used chip area. The largest portion is attributed to the worker cores with their memories and, to some extent, the DDR4 memory controller. Specifically, the worker cores and their scratchpads are the dominant consumers of DSP and BRAM cells.

%
%

\section{Conclusion}\label{sec:conclusion}
High-performance hardware providing real-time guarantees is required for real-time systems employing neural networks.  For this application scenario, we have developed and implemented a multi-core vector processor, that is be published under a open-source license. This processor utilizes local memories, allowing the processor cores to operate independently, and a central management core with DMA which orchestrates the data transfers between local memories and a main memory following a static schedule.

The evaluation shows that the multi-core architecture provides a performance increase compared to a single large vector processor despite the communication overhead between local memories. At the same time, time-predictability is maintained, as the architecture prevents memory interference by design.

Furthermore, the multi-core platforms achieve higher clock frequencies due to shorter critical paths, yielding an additional performance benefit. However, scalability limits are encountered at 16 cores due to congestion in the FPGA routers. 

While the multi-core platform requires slightly more FPGA resources compared to the single-core baseline, it nonetheless demonstrates good scalability in terms of resource utilization.

%
%
%
\bibliographystyle{splncs04}
\bibliography{literature}

\end{document}